# Saath: Speeding up CoFlows by Exploiting the Spatial Dimension

Akshay Jajoo

Y. Charlie Hu

Rohan Gandhi

Cheng-Kok Koh

## Abstract

CoFlow scheduling improves data-intensive application performance by improving their networking performance. State-of-the-art CoFlow schedulers in essence approximate the classic online Shortest-Job-First (SJF) scheduling, designed for a single CPU, in a distributed setting, with no coordination among how the flows of a CoFlow at individual ports are scheduled, and as a result suffer two performance drawbacks: (1) The flows of a CoFlow may suffer the out-of-sync problem – they may be scheduled at different times and become drifting apart, negatively affecting the CoFlow completion time (CCT); (2) FIFO scheduling of flows at each port bears no notion of SJF, leading to suboptimal CCT.

We propose SAATH, an online CoFlow scheduler that overcomes the above drawbacks by explicitly exploiting the spatial dimension of CoFlows. In SAATH, the global scheduler schedules the flows of a CoFlow using an *all-or-none* policy which mitigates the out-of-sync problem. To order the CoFlows within each queue, SAATH resorts to a Least-Contention-First (LCoF) policy which we show extends the gist of SJF to the spatial dimension, complemented with starvation freedom. Our evaluation using an Azure testbed and simulations of two production cluster traces show that compared to Aalo, SAATH reduces the CCT in median (P90) cases by 1.53× (4.5×) and 1.42× (37×), respectively.

## 1 Introduction

In analytics at scale, speeding up the communication stage directly helps to speed up the analytics jobs. In such settings, network-level metrics such as flow completion time (FCT) do not necessarily improve application-level metrics such as job completion time [16, 17, 22]. The CoFlow abstraction [16] is proposed to capture the network requirements of data-intensive applications so that improving network-level performance directly improves application-level performance.

In particular, a CoFlow consists of multiple concurrent flows within an application that are semantically synchronized; the application cannot make progress until all flows in a CoFlow have completed. Since in compute clusters, each job may consist of one or more CoFlows, and multiple jobs share the network fabric, it raises the *CoFlow scheduling problem* with the objective of minimizing the overall CoFlow Completion Time (CCT) (NP-hard [19, 40]).

State-of-the-art CoFlow schedulers such as Aalo [17] in essence apply the classic online approximate Shortest-Job-First (SJF) algorithm using priority queues, where shorter CoFlows finish in high priority queues, and longer CoFlows do not finish in high priority queues, and are moved to and will finish in low priority queues.

Since a CoFlow has many flows distributed at many network ports, Aalo approximates the online SJF, designed for a single CPU, in a distributed setting. It uses a global coordinator to sort the CoFlows to the logical priority queues based on the progress made (total bytes sent); the flows of a CoFlow are assigned to the same priority queue at all network ports. At each port, the local scheduler applies a FIFO policy to schedule flows in each priority queue.

We make a key observation that this way of dividing the CoFlow scheduling task fundamentally does not take into account the *spatial dimension* of CoFlows scheduling, *i.e., once assigned to the priority queues, the individual flows of a CoFlow are scheduled without any coordination until the CoFlow switches the queue or its flows finish.* Such lack of coordination in turn leads to two problems that negatively impact the quality of the scheduling algorithm.

**Out-of-sync problem:** First, the flows of a CoFlow at different ports can get scheduled at different times, which we refer to as the *out-of-sync* problem. Since the CCT is determined by the flow that completes the last, the flows that completed earlier did not help the CCT, but unnecessarily blocked or delayed the flows of some other CoFlows in their respective local ports, affecting the CCT of those CoFlows. Our evaluation using a production cluster trace shows that the out-of-sync problem is prevalent and severe (§2.3): over 20% of CoFlows with equal-length flows experience over 39% normalized deviation in FCT.

**Contention-Oblivion problem:** Second, when taking into account the spatial dimension of CoFlows, we observe that SJF (based on the total bytes of CoFlows) is not optimal in the first place. Intuitively, in a single-CPU job scheduling of $N$ jobs, scheduling any job to run first will block the same number of other jobs, $N − 1$. In scheduling CoFlows across ports, however, since different CoFlows have different numbers of flows distributed at the ports, scheduling a different

CoFlow (its flows) first can block a different number of other CoFlows (at the ports where its flows lie). We denote this degree of competition as *CoFlow contention*. In other words, the waiting time of other CoFlows will depend on the duration as well as the *contention* of the CoFlow across its ports. Both SJF and Smallest Effective Bottleneck First (SEBF) [19] only consider CoFlow duration, and ignore the contention which can result in poor CCT.

In this paper, we propose a new online CoFlow scheduling algorithm SAATH.[1] Like Aalo, SAATH is an online CoFlow scheduler that does not require apriori knowledge of CoFlows. Unlike Aalo, SAATH explicitly takes into account the spatial dimension in scheduling CoFlows to overcome the out-of-sync and contention-oblivion drawbacks of prior CoFlow scheduling algorithms. SAATH employs three key ideas. First, it mitigates the out-of-sync problem by scheduling CoFlows using an *all-or-none* policy, where all the flows of a CoFlow are scheduled simultaneously. Second, to decide on which CoFlow to schedule first following all-or-none, SAATH implements contention-aware CoFlow scheduling. As the CoFlow durations are not known apriori, SAATH adopts the same priority queue structure as Aalo and starts all CoFlows from the highest priority queue on their arrival. Instead of FIFO, SAATH schedules CoFlows from the same queue using *Least Contention First* (LCoF), where the contention due to one CoFlow is computed as the number of other CoFlows blocked on its ports when the CoFlow is scheduled. LCoF prioritizes the CoFlows of less contention to reduce the total waiting time. SAATH further uses CoFlow deadlines to avoid starvation. In contrast, Aalo [17] uses FIFO for online CoFlow scheduling, and other scheduling policies in Varys [19] (including SEBF) are offline and require apriori knowledge about CoFlow sizes.

Third, we observe that using the total bytes sent to sort CoFlows to priority queues ignores the spatial dimension and worsens the out-of-sync problem. When some flows of a CoFlow are scheduled due to out-of-sync, that CoFlow will take *longer* to reach the total-bytes queue threshold, which leads to other CoFlows being blocked on those ports for longer durations, worsening their CCT. SAATH addresses this problem by using a *per-flow* queue threshold, where when at least one flow crosses its share of the queue threshold, the entire CoFlow moves to the next lower priority queue.

Additionally, SAATH handles several practical challenges in scheduling CoFlows in compute clusters in the presence of dynamics, including stragglers, skew and failures.

We implemented and evaluated SAATH using a 150-node prototype deployed on a testbed in Microsoft Azure, and large-scale simulations using two traces from production clusters. Our evaluation shows that, in simulation, compared to Aalo, SAATH reduces the CCT in median case by 1.53× and 1.42× (P90 = 4.5× and 37×) for the two traces while avoiding starvation. Importantly, this CCT reduction translates into a reduction in the job completion time in testbed experiments by 1.46× on average (P90 = 1.86×).

---

[1] SAATH implies a sense of togetherness in Hindi.

In summary, this paper makes the following contributions:
- Using a production datacenter trace from Facebook, we show the prevalence of the *out-of-sync* problem in existing CoFlow scheduler Aalo, where over 20% of CoFlows with equal-length flows experience over 39% normalized deviation in FCT.
- We show that the SJF (and also Shortest-Remaining-Time-First) scheduling policies are not optimal in CoFlow scheduling as they ignore contention across the parallel ports when scheduling CoFlows.
- We present the design, implementation and evaluation of SAATH that explicitly exploits the *spatial dimension* of CoFlows to address the limitations of the prior art, and show the new design reduces the median (P90) CCT by 1.53× (4.5×) and 1.42× (37×) for two production cluster traces.

## 2 Background

In this section, we provide a brief background on the CoFlow abstraction, the Aalo scheduler, and its limitations.

### 2.1 CoFlow Abstraction

In data-intensive applications such as Hadoop [1] and Spark [3], the job completion time heavily depends on the completion time of the communication stage [12, 18]. The CoFlow abstraction was proposed to speed up the communication stage [16] to improve application performance. A CoFlow is defined as a set of flows between several nodes that accomplish a common task. For example, in map-reduce jobs, typically a CoFlow is a set of all flows from all map to all reduce tasks in a single job. The CoFlow Completion Time (CCT) is defined as the time duration between when the first flow arrives and the last flow completes. In such applications, improving CCT is more important than improving the individual flows completion times (FCTs) to improve the application performance [17, 19, 22, 28].

### 2.2 Aalo Scheduler

A classic way to reduce the overall CCT is SCF (Shortest CoFlow First) [17, 19] derived from classic SJF (Shortest Job First). However, using SCF *online* is not practical as it requires prior knowledge about the CoFlow sizes. This is further complicated as CoFlows arrive and exit dynamically and by cluster dynamics such as failures and stragglers.

Aalo [17] was proposed to schedule CoFlows online without any prior knowledge. Aalo approximates SCF using: (1) discrete priority queues, and (2) transitioning the CoFlows across the queues using the total bytes sent so far by a CoFlow. In particular, Aalo starts all CoFlows in the highest priority queue and gradually moves them to the lower priority queue as the CoFlows send more data and exceed the per-queue thresholds. This design choice facilitates the completion of shorter CoFlows as known longer CoFlows move to lower priority queues, making room for potentially shorter CoFlows in the higher priority queues.

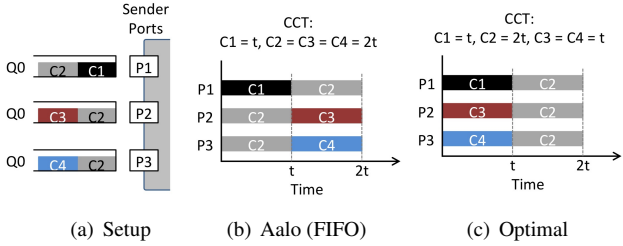

**Figure 1: The out-of-sync problem in SAATH.** The arrival time for CoFlow $C_1 < C_2 < C_3 < C_4$. The individual CCTs in Aalo (average=1.75·t) and optimal case (average=1.25·t) are denoted in Fig (b) and (c).

To implement the above online approximate SCF in a distributed setting, Aalo uses a global coordinator to assign CoFlows to logical priority queues. At each network port, the individual local ports then act *independently* in scheduling flows in its local priority queues, *e.g.,* by enumerating flows from the highest to lowest priority queues and using FIFO to order the flows in the same queue. In doing so, Aalo is oblivious to the *spatial dimension*, *i.e.,* it does not coordinate the flows of a CoFlow across different ports, which leads to two performance drawbacks.

### 2.3 Drawback 1: Out-of-Sync Problem

As individual ports locally have flows of different CoFlows, FIFO can result in the *out-of-sync problem*, *i.e.,* flows of a CoFlow are scheduled at different times at different ports, as shown in the example in Fig. 1. The out-of-sync problem can substantially worsen the overall CCT in two ways:

(1) Since the CCT depends on the completion time of the bottleneck (slowest) flow of the CoFlow, even if non-bottleneck flows of a CoFlow finish earlier, doing so does not improve the CCT of that CoFlow. Instead, such scheduling could block other potentially shorter CoFlows at those ports, and hence worsen their CCT (Fig. 1).

(2) Aalo uses the *total bytes* sent so far to move CoFlows down the priority queues which further worsens the above problem. When only a subset of the flows of a CoFlow are scheduled, it would take longer to reach the same total-bytes queue-crossing threshold compared to when all the flows are scheduled. Hence, the scheduled flows occupy their ports for longer time, which does not improve their CCT, yet may worsen the CCT of other CoFlows that otherwise could have been scheduled.

To understand the extent of the out-of-sync problem, we analyze the variance of the flow completion time of each CoFlow under Aalo, using a trace from Facebook clusters [4]. First, Fig. 2(a) plots the distribution of the number of flows per CoFlow, and Fig. 2(b) plots the distribution of the standard deviation of flow lengths per CoFlow, normalized by its average flow length. We see that in the FB trace, 23% of the CoFlows have a single flow, 50% have multiple, equal-length flows, and the remaining 27% have multiple, unequal-length flows. We then plot the standard deviation of FCT of each of the multi-flow CoFlows, normalized by the average FCT of its flows. We note that the flows of a CoFlow can be of uneven length, which can contribute to uneven FCT. To isolate this factor, in Fig. 2(c), we separately show this distribution for CoFlows with equal and unequal flow lengths (excluding single-flow CoFlows). We see that the out-of-sync problem under Aalo is severe: the FCT of 50% (20%) of the equal-flow-length CoFlows have over 12% (39%) normalized deviation, and of the CoFlows with multiple, uneven-length flows, 50% (20%) have over 27% (50%) normalized deviation in FCT.

### 2.4 Drawback 2: SJF is Sub-optimal for CoFlows

Assuming that the flows of each CoFlow are now scheduled in synchrony, the coordinator still needs to decide which CoFlows should go first to reduce the overall CCT. SCF derived from SJF has been a de-facto policy [17, 22]. We observe that SCF based on the total bytes sent by CoFlows is not optimal in CoFlow scheduling even in the (ideal) offline settings when the CoFlow sizes are known apriori. Similarly, even the Shortest-Remaining-Time-First (SRTF) which improves SJF by allowing preemption is not optimal even when CoFlow sizes are known apriori. The key reason is that these scheduling policies are designed for scheduling jobs serially on a single work engine. They are oblivious to the spatial dimension of CoFlows, *i.e.,* different flows of a CoFlow may be scheduled concurrently and contend with different numbers of other CoFlows (empirically proven in Appendix). Intuitively, two CoFlows C1 and C2 with durations $t_1$ and $t_2$ may block $k_1$ and $k_2$ other CoFlows when their flows are scheduled across individual ports. For example, in Fig. 1, $k_1$=1, $k_2$=3, $k_3$=$k_4$=1. Thus, the increase of the total waiting time of other CoFlows when scheduling C1 and C2 would be $t_1 \cdot k_1$ and $t_2 \cdot k_2$, respectively. SJF and SRTF only consider $t_1$ and $t_2$, and miss out the $k_1$ and $k_2$ factors, which can result in higher total waiting time for the rest of CoFlows and thus sub-optimal CCT.

As a quick evidence that SJF is not optimal for CoFlow scheduling, we compare it with a *Least-Waiting-Time-First*

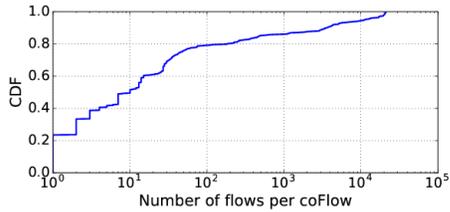

(a) Distribution of CoFlow width

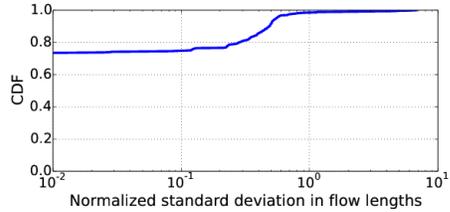

(b) Normalized standard deviation of flow lengths

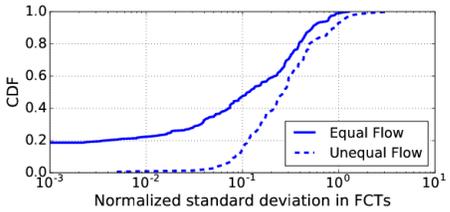

(c) Normalized standard deviation in FCTs for Aalo

Figure 2: The out-of-sync problem in Aalo. (a) Distribution of number of flows in a CoFlow. (b) Distribution of standard deviation of flow lengths normalized by the average flow length, per CoFlow. (c) Distribution of normalized standard deviation of FCTs for multi-flow CoFlows under Aalo. In (c), we have excluded the CoFlows with single flows (23%).

*(LWTF)* policy. In LWTF, the CoFlows are sorted based on the increase in the total waiting time of other CoFlows, *i.e.*, $t \cdot k$. We then compare the improvement of the CCT of individual CoFlows as well as the overall CCT under LWTF, SCF and SRTF over Aalo in the ideal offline settings where the CoFlow sizes are known, using the FB trace. Fig. 3 shows LWTF outperforms SRTF and SCF, suggesting SCF and SRTF are not optimal, and considering *contention* when scheduling CoFlows leads to better CCT.

## 3 Key Ideas

To address the two limitations of Aalo, we propose a new online CoFlow scheduler called SAATH that explicitly takes into account the *spatial dimension* of CoFlows, *i.e.*, the flows of each CoFlow across different network ports. Specifically, SAATH directly tackles the two limitations of Aalo: (1) the out-of-sync problem is mitigated by scheduling all flows of a CoFlow together; (2) the *contention among CoFlows across*

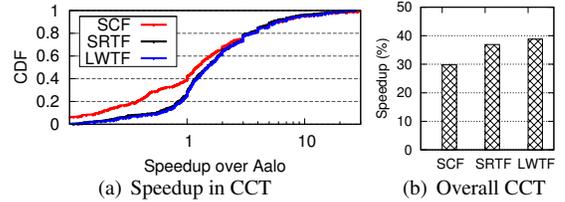

(a) Speedup in CCT  (b) Overall CCT

Figure 3: Comparing CCT speedup using SCF, SRTF and LWTF over Aalo assuming flow statistics are known. The curves for SRTF and LWTF overlap in (a). Overall CCT in (b) is the average CCT for all CoFlows.

*the ports* is explicitly considered in scheduling CoFlows. In the following, we detail on these core ideas that shape the SAATH design.

**(1) All-or-none:** The first key idea in SAATH is to schedule the CoFlows using an *all-or-none* policy, *i.e.*, either all the flows of a CoFlow are scheduled together, or none. This design choice effectively alleviates the out-of-sync problem in Aalo, as the ports that used to schedule a subset of flows of a CoFlow early can now delay scheduling them, without potentially inflating the CCT of that CoFlow, since its CCT depends on the completion of its last flow. The scheduling slots at those ports can be used for some other CoFlows, potentially improving their CCT.

Our key insight is that, in the context of conventional flow scheduling, typically the FCT of one flow cannot be improved without degrading the FCT of another flow [26]. However, this is not true in the context of CoFlows, as the CCT of a CoFlow comprising of many flows depends on the completion of the last flow, and thus a delay in the earlier finishing flows of a CoFlow should not inflate its CCT but could improve the CCT of other CoFlows. In doing so, the CCT of one CoFlow can be improved without worsening the CCT of other CoFlows.

However, all-or-none alone can potentially result in poor port utilization because it requires *all* ports of a CoFlow to be available when scheduling; if not all ports needed by a CoFlow are available, they may be all sitting idle as shown in the example in Fig. 4(b). SAATH carefully designs the work conservation scheme to schedule additional flows at ports that are otherwise left idle, as shown in Fig. 4(c) (§4.2).

One may observe that, as shown in Fig. 4(c), applying work conservation appears to break away from all-or-none. We argue that it does not re-create the out-of-sync limitation in Aalo. Recall that the out-of-sync limitation in Aalo was caused due to scheduling a CoFlow at a time slot that otherwise could have been used for a potentially shorter CoFlow. In SAATH, work-conservation schedules a CoFlow in an otherwise *empty* time slot, which does not push back other CoFlows. Instead it will only speed up the CoFlows.

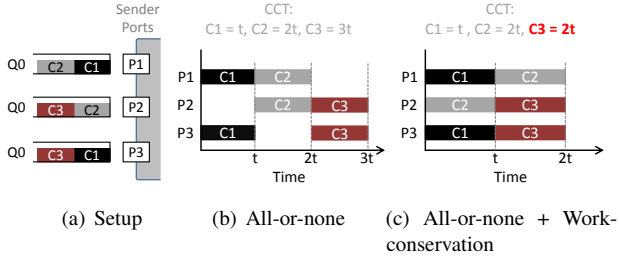

(a) Setup  (b) All-or-none  (c) All-or-none + Work-conservation

**Figure 4:** Unused ports in all-or-none can elongate CCT as in (b), with average CCT = 2·t. (c) Work-conservation can speedup CoFlows (average CCT = 1.67·t).

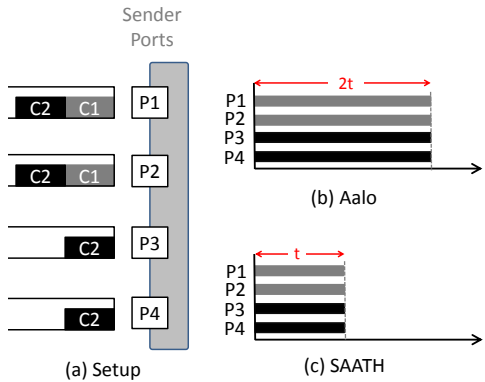

(a) Setup  (b) Aalo  (c) SAATH

**Figure 5:** Fast queue transition in SAATH. (a) CoFlow organization. (b) Transition for C1 and C2 in Aalo. Assume the queue threshold is $bandwidth \cdot 4t$. C2 takes $2t$ time units to reach the threshold as 2 ports (out of 4) are sending data. (c) Fast queue transition in SAATH. The per-flow queue threshold for C2 is $bandwidth \cdot t$ as there are 4 flows, and it takes $t$ time units to reach the threshold.

We note that if the flow lengths in a CoFlow are skewed, all-or-none may not finish all the flows of a CoFlow together. Since SAATH is an online CoFlow scheduler, it does not know the flow lengths beforehand. As a result, in some cases, it may end up delaying scheduling a longer flow to align with other flows, which may delay completion of that flow and worsen the CCT of the CoFlow. Our evaluation (§6.2) shows that such cases are rare, and overall all-or-none improves CCT.

**(2) Faster CoFlow-queue transition:** Since the flow durations are not known apriori, like Aalo, SAATH uses the priority queue structure to approximate the general notion of Shortest CoFlow First, by helping shorter CoFlows finish (early) in high priority queues.

The key challenge in priority queue-based design is to quickly determine the right queue of the CoFlow, so that the time that longer CoFlows contend with the shorter CoFlows is minimized. Like Aalo, SAATH starts all the CoFlows from the highest priority queue. Unlike Aalo, SAATH uses *per-flow queue thresholds*. When an individual flow of a CoFlow reaches its fair share of the queue threshold before others, *e.g.*, from work conservation, we move the *entire CoFlow* to the next lower priority queue.

In essence, if a CoFlow is expected to cross the queue threshold, using per-flow queue thresholds effectively speeds up such queue transition as shown in Fig. 5, where a CoFlow is transitioned to the next queue in time $t$ instead of $2 \cdot t$ as in Aalo. Such faster queue transition has an immediate benefit: it frees the ports where the remaining flows of that CoFlow are falling behind *sooner*, *i.e.*, by moving them to the next lower priority queues at their corresponding ports, so that other high priority CoFlows could be scheduled sooner, potentially improving their CCT.

In SAATH, we calculate the fair share threshold by simply splitting the queue threshold equally among all the flows of a CoFlow. More sophisticated ways can be used in clusters with skewed flow duration distribution.

**(3) Least-Contention-First policy within a queue:** Once the CoFlows are assigned to the priority queues, the next challenge is to order and schedule the CoFlows from the same queue. In SAATH, we propose the Least-Contention-First (LCoF) policy, where the *contention* of a CoFlow is calculated as the number of other CoFlows blocked when that CoFlow is scheduled at all of its ports.[2] Under LCoF, all the CoFlows in each queue are sorted according to the increasing order of contention, and the scheduler scans the sorted list from each queue, starting from the highest priority queue, and schedules the CoFlow that competes against the least number of other CoFlows, as long as there is enough port bandwidth remaining. In essence, by scheduling CoFlows in the LCoF manner, SAATH allows more CoFlows (who have less contention) to be scheduled in parallel in conforming to all-or-none and hence more CoFlows to finish earlier. Our evaluation results (Fig. 10) confirm that SAATH gains significant improvement by use of LCoF.

In summary, SAATH improves the CCT of the CoFlows from the same priority queue using LCoF and all-or-none, and accelerates the CoFlow queue transition using per-flow queue thresholds to further improve the overall CCT.

## 4  Online Scheduler Design

In addition to the three key ideas for improving CCT, SAATH also needs to (1) provide starvation-free guarantee for continuous progress, as LCoF can indefinitely delay scheduling a CoFlow that always has higher contention than other CoFlows,

---

[2]We note the contention thus defined is an approximation to the impact scheduling that CoFlow does to the overall CCT, which should be weighted by the remaining flow lengths of the CoFlow, which however is not known.

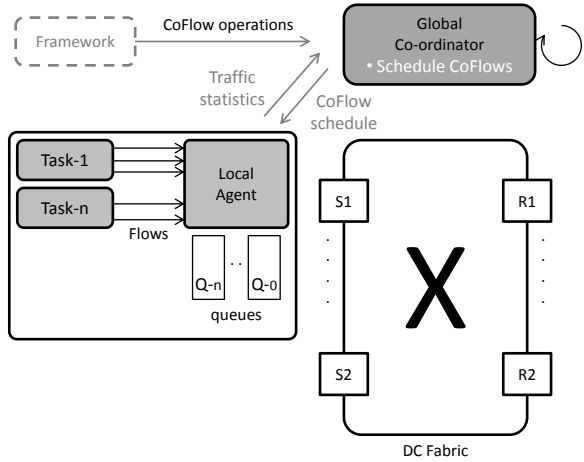

**Figure 6:** SAATH **architecture.**

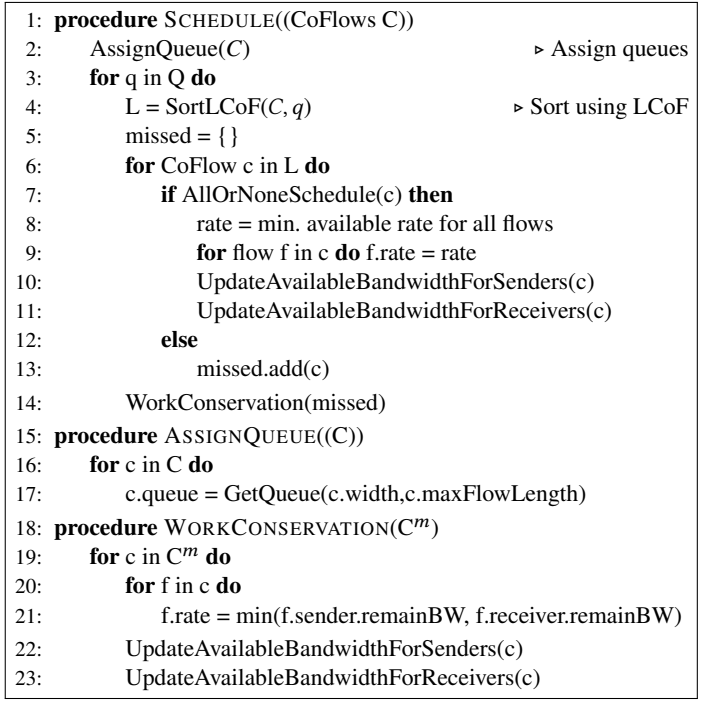

**Figure 7:** SAATH **scheduling algorithm.**

and (2) speed up CoFlows during cluster dynamics such as node failures and stragglers.

In this section, we present the detailed SAATH design to overcome these challenges. The key design features in SAATH are summarized as follows:

(1) *All-or-none*: mitigates the out-of-sync problem;
(2) *Per-flow queue threshold*: speeds up queue transition;
(3) *LCoF*: orders CoFlows within a queue in a contention-aware manner;
(4) *Work-conservation*: improves port utilization and the overall CCT;
(5) *Handling cluster dynamics*: speeds up the flows of a CoFlow due to dynamics such as failures and stragglers by moving the CoFlow back to higher priority queues;
(6) *Starvation-free:* provides starvation-free guarantees.

### 4.1 SAATH Architecture

Fig. 6 shows the SAATH architecture. The key components are the global coordinator and local agents running at the individual ports. A computing framework such as Hadoop or Spark first registers (removes) the CoFlows when a job arrives (finishes). At every fixed scheduling interval, the global coordinator computes the schedule for all the ports based on the CoFlow information from the framework and flow statistics sent by the local agents (which update the global coordinator at each scheduling interval, details in §4.2). The coordinator then pushes the schedule back to the local agents. Local agents maintain the priority queues and use them to schedule CoFlows. They continue to follow the current schedule until a new schedule is received from the global coordinator.

SAATH uses the same queue structure as Aalo, and has the same parameter settings. In SAATH, there are $N$ queues, $Q_0$ to $Q_{N-1}$, with each queue having lower queue threshold $Q_q^{lo}$ and higher threshold $Q_q^{hi}$, and $Q_0^{lo} = 0$, $Q_{N-1}^{hi} = \infty$, $Q_{q+1}^{lo} = Q_q^{hi}$. SAATH uses exponentially growing queue thresholds, *i.e.,* $Q_{q+1}^{hi} = E \cdot Q_q^{hi}$.

### 4.2 SAATH Scheduler

Fig. 7 shows the scheduling algorithm used by the global coordinator to periodically compute the schedule to minimize the CCT using all-or-none, per-flow queue threshold, and LCoF, and to provide starvation-free guarantee.

**Input:** The input to the algorithm includes (1) the set of CoFlows ($C$), (2) traffic sent by the longest flow of every CoFlow ($t_{c,f}$), (3) starvation-free deadline ($d_c$), (4) the ports used by individual CoFlows ($p_{c,p}$), (5) Total capacity (bandwidth) available at $p^{th}$ port ($B_p$).

We calculate $k_c$, the number of CoFlows *contending* with the c-th CoFlow across all the ports. This is used in implementing the LCoF policy.

**Output:** *f.rate*, *i.e.,* the bandwidth assigned to each CoFlow at each port.

**Objective:** Minimize the average CCT.

**D1. Overall algorithm:** (1) First, the coordinator determines the queue of the CoFlows based on the maximum data sent by any flow of a CoFlow, *i.e.,* $m_c = max(\forall_{f \in f_c}, t_{c,f})$ and per-flow threshold (see D3, D4) (line 2). (2) Next, it sorts the CoFlows, starting from the highest priority queue to the lower priority queues. (3) Within each queue, it sorts the CoFlows using LCoF, *i.e.,* based on their $k_c$ values (line 3:4). (4) It then scrolls through CoFlows one by one, and if all the ports of a

CoFlow (sender and receiver) have available bandwidth (line 7), the CoFlow is scheduled. SAATH assigns the bandwidth as discussed in D2 below, based on which the port allocated bandwidth is incremented (line 9, 10). If any of the ports are un-available, the coordinator skips that CoFlow and moves to the next CoFlow. (5) The algorithm terminates when all CoFlows are scanned or all bandwidth is exhausted by work conservation (see D4 below).

**D2. Assigning flow bandwidth:** As in MADD [19], SAATH assigns equal rates (bandwidth) at the ports as there is no benefit in speeding-up flows at certain ports when the CCT depends on the slowest flow. At a port, we use max-min fairness to schedule the individual flows of a CoFlow (to different receivers). Hence, the rate of the slowest flow is assigned to all the flows in the CoFlow, and the port-allocated bandwidths at the coordinator are incremented accordingly.

**D3. Determining CoFlow queue:** Similar to Aalo, SAATH uses exponentially growing queue thresholds. To realize faster queue transition, we divide the queue threshold ($Q_q^{hi}$) equally among all the flows (flow count = $N_c$) of a CoFlow. For example, when a queue threshold is 200MB, a CoFlow with 100 flows has a per-flow queue threshold of 2MB. SAATH assigns CoFlow to a queue based on the maximum data sent by any of its flows, using Eq. (1):

$$\frac{Q_{q-1}^{hi}}{N_c} \leq m_c \leq \frac{Q_q^{hi}}{N_c} \quad (1)$$

**D4. Work conservation:** When following the all-or-none policy, it is possible that some of the ports do not have flows scheduled (§3); these ports can be used to schedule CoFlows outside all-or-none, triggering work conservation (line 14, 18-23). In work conservation, the CoFlows are scheduled based on the ordered list of the un-scheduled CoFlows.

**D5. Starvation Avoidance:** Recall that FIFO provides starvation-free guarantee as every flow in a queue is guaranteed forward progress [17]. Such guarantees are not offered by LCoF. To avoid starvation, the coordinator sets a *deadline* for each CoFlow. Importantly, this deadline is derived based on FIFO. Whenever a CoFlow arrives in a queue, a fresh deadline is set for it. For that, the coordinator first generates FIFO ordering at all ports by enumerating all the CoFlows in that queue. If there are $C_q$ CoFlows in the queue, and $t$ is the minimum time a CoFlow needs to spend in the queue based on the queue threshold, the deadline for the new CoFlow for that queue is set to $d \cdot C_q \cdot t$, where $d$ is a constant (d = 2 in our prototype §6). SAATH then prioritizes the CoFlows that reach their deadlines. Essentially, SAATH provides the same deadline guarantee (within a factor of $d$) as a FIFO based scheduler.

### 4.3 Handling Cluster Dynamics

Compute clusters in datacenters frequently undergo a variety of dynamics including node failures and network congestion. Moreover, even individual jobs may experience stragglers and data skew, multiple stages and waves. In this section, we detail on how SAATH adapts to such dynamics to reduce their impact on the CCT.

**Improving tail due to failures, stragglers, skew:** Cluster dynamics such as node failures and stragglers can delay some flows of a CoFlow, which can result in poor CCT as CCT depends on the completion of the last flow. We observe in such cases, some flows of the CoFlow may have already finished. In such cases, we heuristically make use of the flow length of the completed flows to approximate the SRTF policy to potentially speed up such CoFlows, as follows: (1) the coordinator estimates the length of unfinished flows of a CoFlow using the median flow length of its currently finished flows ($f_e$). (2) It estimates the remaining flow lengths for straggling/restarted flows $f_i^{rem} = f_e - f_i$, where $f_i$ is the flow length so far for the i-th unfinished flow. (3) It estimates the remaining time of a CoFlow as $m_c = max(f_i^{rem})$ since the CCT depends on the last flow, and uses $m_c$ to re-assign the CoFlow to a queue using Eq. 1.

The intuition behind the optimization is that, once some of the flows finish, we no longer need to use the priority queue thresholds to estimate flow lengths – we can simply use $m_c$ as above. The benefit of this approximated SRTF policy is that SAATH can move up a CoFlow from *low to high priority queues* when its flows start to finish; as the remaining flows send more data, $f_i$ increases and thus $f_i^{rem}$ decreases. Moving the CoFlow to a higher queue will accelerate its completion, while following SRTF. We note that calculating $f_e$ as the median of the finished flows is a heuristic; more sophisticated schemes such as Cedar [35] can be used to estimate flow lengths, which we leave as future work.

In contrast, Aalo does not move the CoFlow to the higher priority queues even when fewer and fewer flows are pending, because CoFlow is assigned to a queue based on the total bytes sent so far, which only grows as the flows of a CoFlow send more data.

**Scheduling Multi-stage DAG and multiple waves:** Oftentimes, a single analytics query consists of multiple co-dependent stages, where each stage has multiple jobs. Such queries are represented as a Directed Acyclic Graph (DAG) to capture the dependencies (e.g., Hive [2] on Hadoop or Spark). The DAG representation is available before the start of the query while the scheduler builds the query plans. In SAATH, instead of having one CoFlow for every *job* in a stage, we have one CoFlow for every *stage*. This optimization helps SAATH to slow down some of the fast jobs in one stage without affecting the overall completion time, as the completion

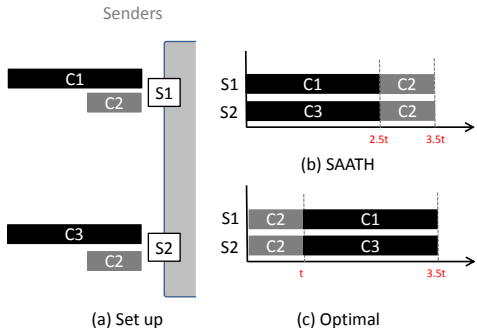

Figure 8: LCoF limitations. (a) shows the setup, CoFlow durations, (b) and (c) show the CoFlow progress in SAATH and optimal. The average CCT in (b) is $\frac{2.5+2.5+3.5}{3} = 2.83$, and in (c) is $\frac{1.0+3.5+3.5}{3} = 2.66$.

of the DAG stage depends on the completion of the slowest job in that stage.

Similarly, a single MapReduce job may have its map and reduce tasks scheduled in multiple waves, where a single wave only has a subset of the map or reduce tasks. We represent such cases again as a DAG, where a single wave is represented as a single CoFlow, and the DAG consists of serialized stages, each with one single CoFlow. In such cases, the goal of DAG scheduling is the same as the CoFlow scheduling, and the same CoFlow scheduling design can be used.

**Un-availability of the data:** Another important challenge is that the data may not always be available in the communication stage as the computing frameworks often *pipeline the compute and communication stages* [45], *i.e.,* the subset of the data is sent from one phase to another as soon as it becomes available, without waiting for the completion of the whole stage. In such frameworks, not all flow data is always available [45] due to some slow or skewed computation. If the coordinator schedules a CoFlow when some of its data is not available, that time slot is wasted.

To address this problem, in SAATH, the ports first accumulate enough data on each of the flows of the CoFlow for one $\delta$, *i.e.,* the interval at which local agents co-ordinate with the coordinator, and explicitly notify the coordinator when such data is available. This information is piggybacked in the flow statistics sent periodically and thus has minimal overhead. The coordinator only schedules the CoFlows that have enough data to send.

### 4.4 LCoF Limitation

Although LCoF substantially outperforms other scheduling policies (§6, §7), there are rare cases where LCoF performs worse. The key reason is that LCoF schedules CoFlows based on the contention; if there are CoFlows that have less contention but are longer in size, scheduling such CoFlows using LCoF would be sub-optimal as shown by the example in Fig. 8. However, our trace shows that such CoFlows only constitute a minor fraction of the total CoFlows (§6.2, bin-2 in Fig. 11 and Fig. 12), and hence their impact is dwarfed by the improvements on other CoFlows from using LCoF.

## 5 Implementation

We implemented SAATH consisting of the global coordinator and local agents (Fig. 6) in 5.2 KLoC in C++.

**Coordinator:** The coordinator schedules the CoFlows based on the operations received from the framework and traffic statistics from the local agents. The key implementation challenge for the coordinator is that it needs to be fast in computing and updating the schedules. The SAATH coordinator is multi-threaded and is optimized for speed using a variety of techniques including pipelining, process affinity, and concurrency whenever possible.

Conceptually, the coordinator computes new schedules in fixed intervals *e.g.,* the time required to send 1MB at a port, which is 8ms with our setting. In practice, due to the delay in computing and propagating the schedules, the coordinator and local agents work in a pipelined manner. In each interval, the coordinator computes a new schedule consisting of the CoFlow order and flow rates, based on the flow stats received during the previous interval, and pushes them to local agents right away. How local agents react is described below.

Since the coordinator makes scheduling decisions on the latest flow stats received from the local agents, it is stateless, which makes it easy for the coordinator to recover from failures. When the coordinator fails, new deadlines are calculated for each CoFlow.

**Local agents:** Upon receiving a new schedule from the coordinator, each local agent schedules the flows accordingly, *i.e.,* they comply to the previous schedule until a new schedule is received. In addition, the local agents periodically, at the same frequency at which the coordinator calculates new schedules, send the relevant CoFlow statistics, including per-flow bytes sent so far and which flows finished in this interval, to the coordinator. To intercept the packets from the flows, local agents require the compute frameworks to replace `datasend()`, `datarecv()` APIs with the corresponding SAATH APIs, which incurs very small overhead. Lastly, the local agents are optimized for low CPU and memory overhead (evaluated in §7.3), enabling them to fit well in the cloud settings [37].

**CoFlow operations:** The global coordinator runs independently from, and is not coupled to, any compute framework, which makes it general enough to be used with any framework. It provides RESTful APIs to the frameworks for CoFlow operations: (a) `register()` for registering a new CoFlow when it enters, (b) `deregister()` for removing a CoFlow when it exits, and (c) `update()` for updating CoFlow status whenever there is a change in the CoFlow structure, particularly during task migration and restarts after node failures.

# 6 Simulation

We evaluated SAATH using a 150-node testbed cluster in Azure that replays the Hive/MapReduce trace from Facebook (FB). In addition, we evaluate SAATH using large-scale simulations using traces from production clusters of Facebook and a large online service provider (OSP). The FB trace is for 150 ports and is publicly available at [4]. The OSP trace is from a Microsoft cluster and has O(1000) jobs collected from O(100) ports.[3] The highlights of these evaluation are:

- SAATH significantly improves the overall CCT. In simulation using the FB trace, the CCT is improved by 1.53× in the median case (P90 = 4.50×). For the OSP trace, the improvements in CCT are 1.42× in the median case (P90 = 37.2×).
- In testbed experiments, compared to Aalo, SAATH improves job completion time by 1.46× on average (P90 = 1.86×);
- The SAATH prototype is fast and has small memory and CPU footprints;
- The breakdown of the improvement justifies the effectiveness of LCoF, all-or-none, and faster queue transition design ideas.

We present detailed simulation results in this section, and the testbed evaluation of our prototype in §7.

**Setup:** In replaying the traces (FB and OSP), we maintain the same flow lengths and flow ports. The *default parameters* in the experiments are: starting queue threshold ($Q_0^{hi}$) is 10MB, exponential growth factor ($E$) is 10, the number of queues ($K$) is set to 10, and the new schedule calculation interval $\delta$ is set to 8ms. Our simulated cluster uses the same number of nodes (network ports) and link capacities as per the trace. We assume full bisection bandwidth supporting 1 Gbps/port is available, and congestion can happen only at the network ports. The simulator is written in 4 KLOC in C++.

We compare SAATH against two start-of-the-art CoFlow schedulers, Aalo and Varys, which are open-sourced [6]. All the experiments use the above default parameters including $K, E, S$, unless otherwise stated. Since Varys does not use multiple queues, there is no use of queueing parameters for Varys related experiments.

## 6.1 CCT Improvements

We first compare the speedup of SAATH over other scheduling policies. We define the *speedup* using SAATH as the ratio of the CCT under other policy to the CCT under SAATH for individual CoFlows. The results are shown in Fig. 9. The Y-axis denotes the median speedup, and error bars denote the 10-th and 90-th percentile speedups. We show the results for the FB and OSP traces. The key observation is that SAATH improves the CCT over Aalo by 1.53× (median) and 4.5× (P90) for the FB trace, and 1.42× (median) and 37.2× (P90) for the OSP trace. Interestingly, SAATH achieves the speedup close to that of SEBF in Varys [19] even though SEBF runs offline and assumes the CoFlow sizes are known *apriori*, whereas SAATH runs online without apriori CoFlow knowledge.

---
[3]We cannot specify the exact numbers for proprietary reasons, which are also excluded from Fig. 2 in §2.

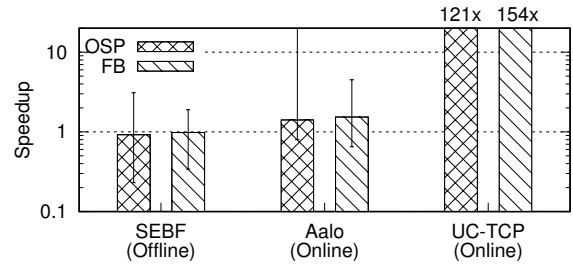

**Figure 9: Speedup using SAATH over other scheduling policies. SAATH achieves speedup of 154× and 121× (median) over UC-TCP for two traces.**

The higher speedup at P90 for the OSP trace over the FB trace is attributed to larger improvement to the CCT of small and narrow CoFlows using all-or-none and LCoF (§6.2). We observe that the ports are busier (*i.e.,* having more CoFlows queued at individual ports) for the OSP trace than the FB trace, which when coupled with FIFO in Aalo, amplifies the waiting time for short and narrow CoFlows in the OSP trace. In contrast, LCoF facilitates such CoFlows, resulting in dramatic reduction in their waiting time.

We also compare SAATH against an un-cordinated CoFlow scheduler (UC-TCP) under which individual ports independently schedule the arriving CoFlows without any global coordinator. In UC-TCP, there are no queues, and all the flows are scheduled upon arrival as per TCP. Lack of coordination, coupled with lack of priority queues severely hampers the CCT in UC-TCP. SAATH achieves a median speedup of 154× and 121× over UC-TCP in the FB and OSP traces, respectively.

These results show that SAATH is effective in accelerating the CoFlows compared to Aalo, and is close in performance compared to Varys which assumes prior knowledge of CoFlow lengths, and achieves high speedups compared to other un-coordinated scheduler.

## 6.2 Impact of Design Components

In this experiment, we evaluate the impact of the individual design components on the speedup in CCT over Aalo. The results are shown in Fig. 10. To better understand the impact, we also show the CCT improvement grouped into different bins based on their width and size of the CoFlows (Table 1) in Fig. 11 and Fig. 12. We make the following key observations.

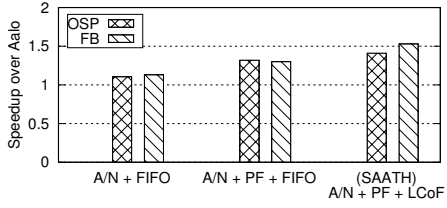

**Figure 10:** SAATH **speedup breakdown across 3 complimentary design ideas. Y-axis shows the median speedup. Abbreviations: (1) A/N: all-or-none, (2) PF: per-flow queue threshold, (3) LCoF: Least-Contention-First.**

**Table 1: Bins based on total CoFlow size and width.**

|  | width ≤ 10 | width > 10 |
|---|---|---|
| size ≤ 100MB | bin-1 | bin-2 |
| size > 100MB | bin-3 | bin-4 |

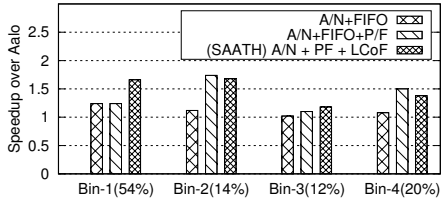

**Figure 11:** SAATH **speedup breakdown into bins based on size and width shown in table 1 for FB trace. The numbers in x-label denote fraction of all CoFlows in that bin. Y-axis shows the median speedup.**

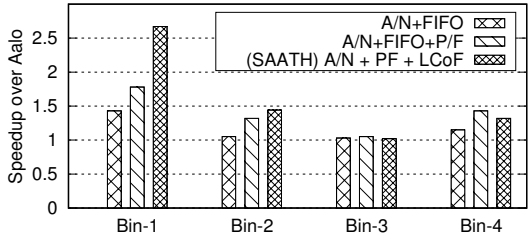

**Figure 12:** SAATH **speedup breakdown into bins based on size and width shown in table 1 for OSP trace. We omit the distribution of CoFlows in individual bins for proprietary reasons. Y-axis shows the median speedup.**

**First,** only using all-or-none (A/N) and FIFO, *i.e.,* without LCoF and per-flow queue threshold (P/F), the speedup over Aalo using FB (OSP) trace is 1.13× (1.1×) in median case and 3.05× (7.2×) at P90. Fig. 11 and Fig. 12 show that all-or-none is effective for small, thin CoFlows, *i.e.,* CoFlows with fewer flows, as the probability of finding all the ports available is higher. For other bins, the benefits are lower due to the use

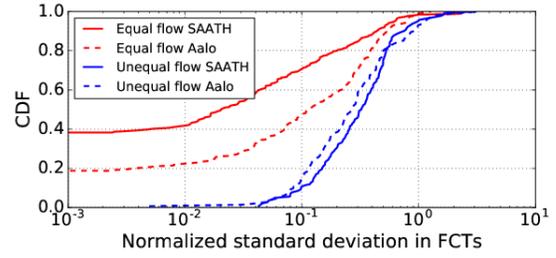

**Figure 13: Normalized standard deviation of FCTs of multi-flow CoFlows, under** SAATH **and Aalo using FB trace. We have excluded the CoFlows with width = 1 (23.5%).**

of FIFO, where a wide CoFlow causes Head-of-Line (HoL) blocking other potentially short CoFlows.

**Second,** while all-or-none only addresses one limitation of the out-of-sync problem, using per-flow queue thresholds (P/F) addresses the second limitation of out-of-sync problem by quickly jumping the queues (§3). As a result, A/N+P/F improves the speedup over Aalo in the FB (OSP) trace to 1.3× (1.32×) in the median case, and 3.83× (13×) at P90.

We again zoom into the improvement in using P/F on CoFlows in different bins, shown in Fig11 and Fig. 12. P/F is highly effective for CoFlows in bins 2 and 4, which are wider (width > 10). The larger numbers of flows in these wider CoFlows increase the chance of at least one flow crossing the per-flow queue threshold and thus move the CoFlows to the next queues faster.

**Third,** we replace FIFO with LCoF and retain A/N and P/F from previous experiment. This combines all the three complimentary ideas in SAATH, and is labeled as SAATH in Fig. 10. We see that using LCoF achieves a median speedup over Aalo of 1.53× (P90=4.5×) for the FB trace, and of 1.42× (P90=37×) for the OSP trace. This is primarily because LCoF schedules CoFlows using Least Contention First and reduces the HoL blocking in FIFO. As shown in Fig. 11 and Fig. 12, LCoF improves the CCT of CoFlows in all bins. Particularly, it substantially benefits *short and thin* CoFlows (bin-1), as HoL blocking due to FIFO blocks these CoFlows the most, without significantly impacting the CoFlows in other bins. This shows that LCoF on top of all-or-none is effective.

Lastly, Fig. 13 shows the CDF of the standard deviation of FCTs of individual CoFlows with more than one flow under SAATH and Aalo for the FB trace. We show results separately for CoFlows with equal and unequal flow length. We see that SAATH significantly reduces the variation in FCTs: 40% of CoFlows with equal flow lengths finished their flows at the same time, as opposed to 20% in Aalo, and 71% of them had normalized FCT deviation under 10%, compared to 47% in Aalo. We note that SAATH does not completely eliminate the out-of-sync problem because of work conservation (§3).

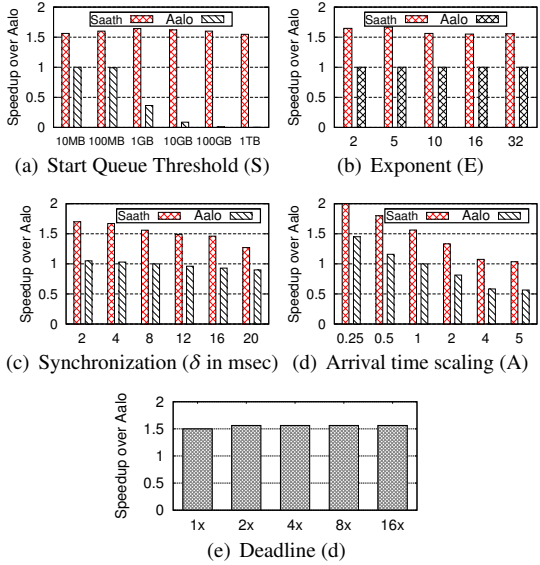

**Figure 14:** SAATH sensitivity analysis.

We do not show the results for the OSP trace for proprietary reasons.

## 6.3 Sensitivity Analysis

We next evaluate the sensitivity of SAATH and Aalo to various design parameters. Due to space limitation, we only show the results for the FB trace. The results for the OSP trace are similar.

**Start queue threshold (S):** In this experiment, we vary queue threshold of the starting (highest priority) queue, which controls how long CoFlows stays in the starting queue. Fig. 14(a) shows that Aalo is highly sensitive to $S$. This is because as $S$ grows, more CoFlows stay in the highest priority queue, and Aalo performs worse due to HoL blocking under FIFO, which is addressed by LCoF in SAATH. In contrast, SAATH is relatively insensitive to $S$, precisely because LCoF alone addresses the HoL blocking weakness of FIFO.

**Multiplication factor (E):** In this experiment, we vary the queue threshold growth factor $E$ from 2 to 32. Recall that the queue thresholds are computed as $Q_q^{hi} = Q_{q-1}^{hi} \cdot E$. Thus, as $E$ grows, the number of queues decreases. As shown in Fig. 14(b), SAATH and Aalo are both insensitive to E.

**Synchronization interval ($\delta$):** Recall that the global coordinator calculates a new schedule every $\delta$ interval. In this experiment, we vary $\delta$ and measure its impact on the CCT. Fig. 14(c) shows as $\delta$ increases, the speedup in Aalo and SAATH both diminish. As shown in §7.3, SAATH comfortably finishes calculating each new schedule within 8 msec even during busy periods (with an average of 0.57 msec and P90 of 2.85 msec). Thus when $\delta$ increases, the CCT increases because the ports may finish the current scheduled flows and become idle before receiving a new schedule from the coordinator. This shows that in general shorter scheduling intervals help to keep all the ports busy which in turn requires the global coordinator to be able to calculate schedules quickly.

**CoFlow arrival time (contention):** In this experiment, we vary the arrival time (A) between the CoFlows to vary the CoFlow *contention*. The x-axis in Fig. 14(d) shows the factor by which the arrival times are sped up. For example, $A = 0.5$ denotes that CoFlows arrive 0.5× faster (2× slower), whereas $A = 4$ denotes that CoFlows arrive 4× faster. The y-axis shows the speedup compared to the default Aalo, *i.e.,* Aalo with $A = 1$. Fig. 14(d) shows that as $A$ increases, the overall speedup in both SAATH and Aalo decreases. This is expected because increasing $A$ causes more contention and the CoFlows are queued up longer increasing their CCT under both schemes.

More importantly, when we increase $A$, the speedup of SAATH over Aalo increases, from 1.53× to 1.9×, showing the higher the contention, the more SAATH outperforms Aalo, using the LCoF policy.

**CoFlow deadline (d):** Recall that LCoF by default does not provide the starvation-free guarantees, and can starve the CoFlows with high contention. To avoid starvation, SAATH assigns each CoFlow a deadline of $d \cdot C_q \cdot t$ (D5 in §4.2), where $C_q \cdot t$ denotes the estimated deadline based on current CoFlows if scheduled under FIFO. In this experiment, we measure the impact of $d$ on the CCT speedup, where $d$ is varied from 1 to 16. Fig. 14(e) shows that SAATH is insensitive to $d$, and comfortably schedules the CoFlows within the deadline. Even when the deadlines are as per FIFO (d = 1), SAATH can achieve a median speedup of 1.5× over Aalo. A small drop in the speedup at d = 1 is because starvation-free scheduling occasionally kicks in, and the CoFlows that passed the deadline are forced to be scheduled, which would not have been under LCoF only, resulting in worse CCT. At higher values of $d$ (> 2), SAATH has more freedom to reorder the CoFlows to facilitate their CCT without violating the deadlines.

## 7 Testbed Evaluation

**Testbed setup:** Similar to simulations, our testbed evaluation keeps the same job arrival times, flow lengths and flow ports in trace replay. All the experiments use the default parameter values of $K, E, S, \delta$. For the testbed experiments, we rerun the trace on Spark-like framework on a 150-node cluster in Microsoft Azure [5]. We use the FB trace as it has a cluster size similar to that of our testbed. The coordinator runs on a Standard F4s VM (4 cores and 8GB memory) based on the 2.4 GHz Intel Xeon® E5-2673 v3 (Haswell) processor. Each local agent runs on a D2v2 VM (2 cores and 7GB memory) based on the same processor and with 1 Gbps network connection.

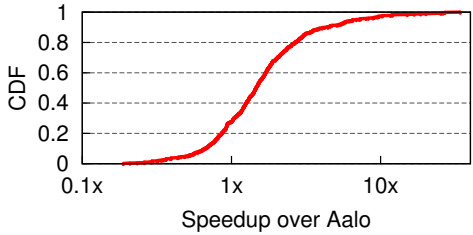

Figure 15: [Testbed] Speedup in CCT in SAATH.

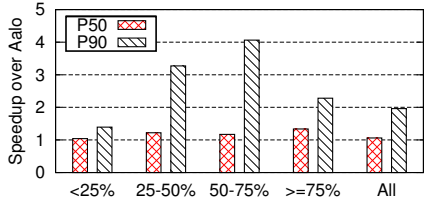

Figure 16: [Testbed] Speedup in job completion time using SAATH over Aalo. X-axis shows the fraction of total job time spent in shuffle phase.

Table 2: [Testbed] Resource usage in SAATH and Aalo.

|  |  | SAATH | | Aalo | |
|---|---|---|---|---|---|
|  |  | Average | P90 | Average | P90 |
| Global co-ordinator | CPU (%) | 37.8 | 42.7 | 33.5 | 35.5 |
|  | Memory (MB) | 229 | 284 | 267 | 374 |
|  | Total time (LCoF / All-or-none) (msec) | 0.57 (0.02 / 0.24) | 2.85 (0.03 / 0.7) | 0.1 | 0.2 |
| Local node | CPU (%) | 5.6 | 5.7 | 5.5 | 5.7 |
|  | Memory (MB) | 1.68 | 1.7 | 1.75 | 1.78 |

In testbed evaluation, we compare SAATH against Aalo. The primary evaluation metric is the speedup in CCT. We also compare the speedup in job completion time and SAATH scheduling overheads.

## 7.1 Improvement in CCT

In this experiment, we measure the speedup in SAATH compared to Aalo. Fig.15 shows that the ratio of CCT under SAATH over that under Aalo ranges between 0.09-12.15×, with an average of 1.88× and a median of 1.43× compared to under Aalo, which is close to the reduction observed in the simulation experiments. Although SAATH improves the CCT for the majority of CoFlows (>70%), it slows down some of the CoFlows. These CoFlows are favored by FIFO as they arrived early. The same CoFlows would be pushed back by SAATH if they observe high contention. Additionally, the starvation avoidance rarely kicked in ($< 1\%$) even for $d = 2$. This experiment shows that SAATH is effective in improving CCT in real settings.

## 7.2 Job Completion Time

Next, we evaluate how the improvements in CCT affects the job completion time. In data clusters, different jobs spend different fractions of their total job time in data shuffle. In this experiment, the fraction of time that the jobs spent in the shuffle phase follows the same distribution used in Aalo [17]. Fig. 16 shows that SAATH substantially speeds up the job completion time of the shuffle-heavy jobs (shuffle fraction ≥50%) by 1.83× on average (P50 = 1.24× and P90 = 2.81×). Additionally, across all jobs, SAATH reduces the job completion time by 1.42× on average (P50 = 1.07× and P90 = 1.98×).

This shows that the benefits in improving CCT translates into better job completion time. As expected, the improvement in job completion time is smaller than the improvements in CCT because job completion time depends on time spent in both compute and communication (shuffle) stages, and SAATH improves only the communication stage.

## 7.3 Scheduling Overhead

We next evaluate the overheads in SAATH at the coordinator and the local agents. Table 2 shows the overheads in terms of CPU and memory utilization for both SAATH and Aalo. We measure the overheads in two cases: (1) Average: the average utilization during the entire execution of the trace, (2) Busy: the 90-th percentile utilization indicating the performance during busy periods when a large number of CoFlows arrive. As shown in Table 2, SAATH has a very small overhead at the local nodes, where the CPU and memory utilization is minimal even during busy times. The global coordinator also uses the server resources economically – compared to Aalo, overall SAATH incurs 4.3% increase in average CPU utilization. Finally, the scheduling latency is overall small, although higher than Aalo due to all-or-none, LCoF and per-flow scheduling. The time it takes the coordinator to calculate new schedules is 0.57 msec on average and 2.85 msec at P90.

We also break down the computation time at the coordinator in SAATH into the time spent in ordering CoFlows (using per-flow thresholds and LCoF), scheduling using all-or-none, and the rest which is for assigning rates for work conservation. Table 2 shows that most of the computation time is spent on assigning rates for work conservation; ordering the CoFlows using LCoF accounts for less than half of the schedule compute time.

In summary, our overhead evaluation shows that the cost of the SAATH scheduling algorithm is moderate, and that the CCT improvement in SAATH outweighs its costs.

## 8 Related Work

**Non-Clairvoyant Scheduling**: Non-clairvoyant scheduling, *i.e.,* scheduling without prior knowledge, has been studied as early as in time-sharing systems [20, 21], with many variations [38, 41] and applied to diverse scheduling problems such as memory controller scheduling [34].

**CoFlow scheduling:** Varys [19] was proposed to schedule the CoFlows assuming *prior* information is available. In contrast, LCoF is an online scheduling policy, that in fact performs comparable to offline SEBF from Varys (Fig. 9). [40] shows CoFlow scheduling is NP hard and proposes heuristics to reduce the average CCT, again assuming prior knowledge of CoFlows. Baraat [22] and Aalo [17] schedule CoFlows in online settings. We have already discussed Aalo extensively. Baraat is a completely de-centralized scheduler without coordination among the ports, and suffers from the same limitation as Aalo. In contrast, SAATH [27, 29] is, to our best knowledge, the first CoFlow scheduler that takes into consideration the spatial dimension when scheduling CoFlows, by applying all-or-none and LCoF. Recently, CODA [48] was proposed to automatically identify flows that belong to the same CoFlows, and schedule them while tolerating identification errors.

**Flow scheduling:** There has been a rich body of prior work on flow scheduling. Efforts to minimize FCTs, both with prior information (*e.g.,* PDQ [26], pFabric [8]) and without prior information (*e.g.,* Fastpass [39], PIAS [13]), fall short in minimizing CCTs which depend on the completion of the last flow [19]. Similarly, Hedera [7] and MicroTE [14] schedule the flows with the goal of reducing the overall FCT, which again is different from reducing the overall CCT of CoFlows. In this [27, 30, 31] paper authors explore a learning technique for coflows.

**All-or-none:** The all-or-none principle in SAATH is also used in different contexts (*e.g.,* caching in Pacman [11], task placement in [25]). Squall [44] uses it to partition data for scalable and fast query processing. Fastpass [39] uses it to schedule flows in datacenters.

**Job scheduling:** There have been much work on scheduling in analytic systems and storage at scale by improving speculative tasks [10, 12, 27, 32, 33, 47], improving locality [9, 46], and end-point flexibility [15, 24, 43]. The CoFlow abstraction is complimentary to these work, and can benefit from them.

**Scheduling in parallel processors:** CoFlow scheduling by exploiting the spatial dimension bears similarity to scheduling processes on parallel processors and multi-cores, where many variations of FIFO [42], FIFO with back-filling [36] and gang scheduling [23] are proposed.

## 9 Conclusion

In this paper, we show that the prior-art CoFlow scheduler Aalo suffers from the out-of-sync problem and by using a simple FIFO scheduling policy which ignores CoFlow contention across ports. We present SAATH that addresses the limitations in Aalo by exploiting the spatial dimension in scheduling CoFlows. SAATH uses all-or-none to schedule all flows of a CoFlow together, and the Least-Contention-First policy to decide on the order of CoFlows from the same priority queue. Our evaluation using a 150-node testbed in Microsoft Azure and large scale simulations using traces from two production clusters shows that compared to Aalo, SAATH reduces the CCT by 1.53× and 1.42× in median, and 4.50× and 37× at the 90-th percentile for two traces.

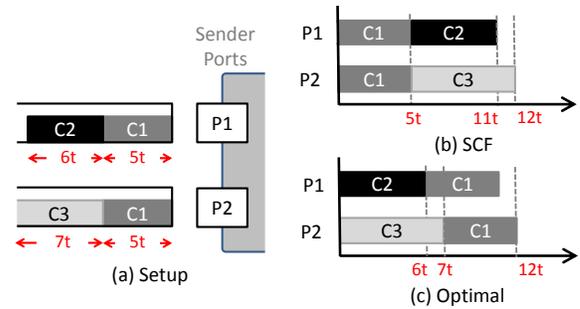

Figure 17: SJF is sub-optimal. (a) shows the setup and CoFlow durations, whereas (b) and (c) show the CoFlow progress. The average CCT in (b) is $\frac{5+11+12}{3} = 9.3$, and in (c) is $\frac{12+6+7}{3} = 8.3$.

## Appendix A: SJF for CoFlows is Sub-optimal

Fig. 17 illustrates that SJF is *not optimal* even when all the CoFlows arrive at the same time, and their durations are known apriori. When an i-th CoFlow is scheduled, the increase in the waiting time of other CoFlows is $t_i \cdot k_i$, where $t_i$ is the duration CoFlow $i$ is scheduled, and $k_i$ is the contention, *i.e.,* the number of other CoFlows blocked. In CoFlows, $k_i$ is non-uniform as different ports have different CoFlows and differen CoFlows reside at different numbers of ports. However, SJF only considers $t_i$, and is agnostic to $k_i$, and thus results in higher waiting time as shown in Fig. 17. In this example, $k_1 = 2, k_2 = k_3 = 1$, and $t_1, t_2, t_3$ are shown in the figure. SJF schedules CoFlow C1 first as it is the shortest. However, C1 blocks the other two CoFlows, increasing the total waiting time by $2 \cdot 5t$, which leads to sub-optimal average CCT.